# Anisotropy of Neel "orange-peel" coupling in magnetic multilayers

M.A. Kuznetsov[1,3], O.G.Udalov[1,2,*], A.A.Fraerman[1,3]

[1]Institute for physics of microstructures RAS, 7 Academicheskaya Str., Afonino, Nizhny Novgorod region, Kstovsky district, Kstovo region, 603087, Russia
[2]California State University Northridge, 18111 Nordhoff Street, Northridge, CA 91330, USA
[3]Lobachevsky State University of Nizhni Novgorod, 23 Prospekt Gagarina (Gagarin Avenue), Nizhny Novgorod, 603950, Russia

* Corresponding author e-mail: oleg.udalov@csun.edu

*Highlights:*
- "Orange-peel" coupling between two ferromagnetic layers is calculated for an arbitrary orientation of the layers magnetizations
- "Orange-peel" coupling changes sign when the magnetizations rotate from in-plane orientation to out-of-plane orientation

*Abstract:* We calculate the energy of the magnetostatic interaction between two ferromagnetic films with uniform magnetization and correlated interfaces (the "orange-peel" effect). The "orange-peel" coupling is anisotropic: the interaction is ferromagnetic when the films are magnetized in-plane; and it is antiferromagnetic when magnetization is out-of-plane. The interaction anisotropy can be used to distinguish the "orange-peel" effect from the interlayer exchange coupling.

The magnetostatic interaction of two magnetic layers appears due to the roughness of their interfaces. The roughness produces a stray field. This field acts on the neighboring film leading to the so-called "orange-peel" interlayer interaction [1-8]. Often the "orange-peel" (OP) coupling is comparable to the interlayer exchange interaction [5, 9-14], which appears due to spin currents of conduction electrons [15]. These two interaction mechanisms should be distinguished in certain cases. This can be done by studying the angular dependencies of the interlayer coupling energy. The exchange interaction is isotropic and depends on the mutual orientation of magnetizations $W_{\text{ex}} \sim (\mathbf{M}_1 \mathbf{M}_2)$, where $\mathbf{M}_{1,2}$ are layers magnetic moments. In the present manuscript we will show that "orange-peel" interaction is anisotropic and obeys the relation $W_{\text{op}} \sim J_\perp M_{1\perp} M_{2\perp} - J_{||} M_{1||} M_{2||}$, where $M_{1,2\perp}$ are the out-of-plane magnetizations components, $M_{1,2||}$ are the in-plane components, $J_\perp$ and $J_{||}$ are positive and of the same order.

Consider a system of two ferromagnetic films with rough surfaces (see Fig. 1). We assume that these surfaces are correlated and can be described by the same function $\xi(\boldsymbol{\rho})$, where $\boldsymbol{\rho} = (x, y)$ is the in-plane coordinate vector. Often the correlation between the film interfaces naturally appears during the films growth [16]. Therefore, this assumption is quite relevant. The ferromagnetic films are separated by a non-magnetic spacer of thickness $a$. We also assume that both layers are uniformly magnetized. Their magnetizations are defined as follows $\mathbf{M}_1 = M_1(\cos(\alpha), 0, \sin(\alpha))$, $\mathbf{M}_2 = M_2(\cos(\beta), 0, \sin(\beta))$. The thicknesses of the layers are $h_1$ and $h_2$.

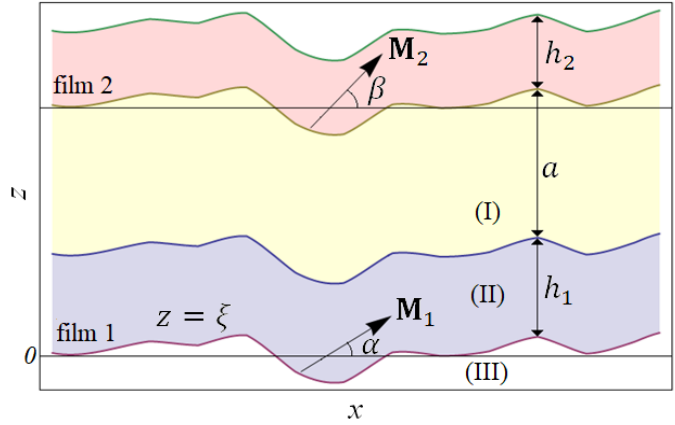

Fig. 1. Two rough ferromagnetic layers separated by a non-magnetic spacer. $h_1$ and $h_2$ denote the thicknesses of the magnetic layers. Spacer thickness is $a$. Horizontal lines show average interfaces. $\mathbf{M}_{1,2}$ are the magnetizations of the ferromagnetic layers. $\alpha$ and $\beta$ are the angles between $\mathbf{M}_{1,2}$ and the average interfaces. $\xi(x, y)$ is the function describing the system roughness.

The magnetostatic energy is given by

$$W_{\text{op}} = -M_2 \int dx dy \left[ \int_{z=h_1+a+\xi(\rho)}^{z=h_1+a+h_2+\xi(\rho)} (\cos(\beta) \cdot H_x + \sin(\beta) \cdot H_z) dz \right], \tag{1}$$

where $H_{x,z}$ are the x- and z-components of the magnetic field produced by the first layer in the region of the second layer. The field can be found using Maxwell equations

$$\text{div}\mathbf{B} = 0, \quad \text{rot}\mathbf{H} = 0. \tag{2}$$

Introducing the magnetic scalar potential $\varphi$ ($\mathbf{H} = -\nabla\varphi$) one can transform Eqs. (2) to

$$\Delta\varphi = 0. \tag{3}$$

At the interfaces $z = h_1 + \xi$ and $z = \xi$ we require that the normal component of the magnetic induction $\mathbf{B}$ and the scalar potential $\varphi$ are continuous

$$-(\mathbf{n}\nabla\varphi^{(I)}) = -(\mathbf{n}\nabla\varphi^{(II)}) + 4\pi(\mathbf{M}_1\mathbf{n})|_{z=h_1+\xi} \tag{4}$$

$$-(\mathbf{n}\nabla\varphi^{(III)}) = -(\mathbf{n}\nabla\varphi^{(II)}) + 4\pi(\mathbf{M}_1\mathbf{n})|_{z=\xi} \tag{5}$$

$$\varphi^{(I)} = \varphi^{(II)}|_{z=h_1+\xi}, \varphi^{(II)} = \varphi^{(III)}|_{z=\xi}. \tag{6}$$

Here $\varphi^{(II)}$ is the potential inside the first ferromagnetic film, notations $\varphi^{(I)}$ and $\varphi^{(III)}$ are used for the potential above ($z > h_1 + \xi$) and below ($z < \xi$) the film, $\mathbf{n}$ is the local normal given by

$$\mathbf{n} = \left(\frac{\xi_x}{\sqrt{1+\xi_x^2+\xi_y^2}}, \frac{\xi_y}{\sqrt{1+\xi_x^2+\xi_y^2}}, -\frac{1}{\sqrt{1+\xi_x^2+\xi_y^2}}\right), \xi_x = \frac{\partial \xi}{\partial x}, \xi_y = \frac{\partial \xi}{\partial y}. \quad (7)$$

We assume that the characteristic roughness height $\xi_0$ is small comparing to the characteristic lengthscale $L$ of the roughness in the (x,y)-plane, $\xi_0 \ll L$. Then we can express the scalar potential as a series

$$\varphi = \varphi_0 + \varphi_1 + \dots, \quad (8)$$

in which $\varphi_n \sim \xi_0^n$ ($n > 1$). The potential $\varphi_0$ is given by

$$\varphi_0 = \begin{cases} 2\pi M_1 h_1 \sin(\alpha), & z > h_1 + \xi, \text{region } (I), \\ 4\pi M_1 (z - \frac{h_1}{2})\sin(\alpha), & \xi < z < h_1 + \xi, \text{region } (II), \\ -2\pi M_1 h_1 \sin(\alpha), & z < \xi, \text{region } (III). \end{cases} \quad (9)$$

Note that $\varphi_0$ is not the solution for the case of a flat surface ($\xi = 0$). We choose the potential $\varphi_0$ pursuing two goals. At first, far from the interface ($z > \xi$) the potential should turn into the solution for the flat interface. The second one is that the $\varphi_0$ should be split into two regions along the true interface. This makes solution of the boundary condition equations much easier. Choosing the zero-order potential as the solution for the perfectly flat surface requires reducing of the boundary conditions to the flat surface as well. While this is also possible, the procedure considered here looks more relevant.

The non-zero magnetostatic interaction energy appears in the second order in $\xi_0$ ($W_{\text{op}} \sim \xi_0^2$), therefore one can calculate the scalar potential up to the first order in $\xi_0$. In the zero order one has

Introducing Eqs. (8) and (9) into Eqs. (3-6) one gets the following equations for $\varphi_1$

$$\varphi_1^{(I)} = \varphi_1^{(II)} + 4\pi M_1 \xi \sin(\alpha)|_{z=h_1}, \varphi_1^{(III)} = \varphi_1^{(II)} + 4\pi M_1 \xi \sin(\alpha)|_{z=0}, \quad (10)$$

$$\frac{\partial \varphi_1^{(I)}}{\partial z} = \frac{\partial \varphi_1^{(II)}}{\partial z} + 4\pi M_1 \xi_x \cos(\alpha)|_{z=h_1}, \quad (11)$$

$$\frac{\partial \varphi_1^{(III)}}{\partial z} = \frac{\partial \varphi_1^{(II)}}{\partial z} + 4\pi M_1 \xi_x \cos(\alpha)|_{z=0}. \quad (12)$$

We introduce the in-plane Fourier components of the potential as follows $\varphi_1(\mathbf{k}, z) = \int \varphi_1(\boldsymbol{\rho}, z) e^{-i\mathbf{k}\boldsymbol{\rho}} dx dy$. The z-dependence of the Fourier components can be expressed in the form

$$\varphi_1(\mathbf{k}) = \begin{cases} A(\mathbf{k})e^{-kz}, & z > h_1 + \xi, \text{region } (I), \\ B(\mathbf{k})e^{-kz} + C(\mathbf{k})e^{kz}, & \xi < z < h_1 + \xi, \text{region } (II), \\ D(\mathbf{k})e^{kz}, & \xi < z, \text{region } (III), \end{cases} \quad (13)$$

where $k = |\mathbf{k}|$ is the absolute value of the wavevector. The coefficients $A, B, C$ and $D$ in Eq. (13) are defined by the boundary conditions Eqs. (10) – (12). Note that the potential $\varphi_1(\mathbf{k})$ should be expanded into a series with respect to $k\xi_0 \sim \xi_0/L$ around the lines $z = h_1$ and $z = 0$ when introducing into Eqs. (10-12). Only the lowest term should be conserved. The terms of higher order in $k\xi_0$ will contribute to the higher order corrections $\varphi_2$, $\varphi_3$ etc. Since we are interested in the potential in the region of the second magnetic film (region (I)) we provide here only the expression for the coefficient $A$

$$A(\mathbf{k}) = 4\pi M_1 e^{\frac{kh_1}{2}} Sh\left(\frac{kh_1}{2}\right) \xi(\mathbf{k}) \left(\sin(\alpha) - i\frac{k_x}{k}\cos(\alpha)\right), \qquad (14)$$

$$\varphi_1^{(I)}(\boldsymbol{\rho}) = \frac{1}{(2\pi)^2} \int A(\mathbf{k}) e^{-kz} e^{i\mathbf{k}\boldsymbol{\rho}} dk_x dk_y \qquad (15)$$

For the interaction energy one has

$$W_{op} = -J_x \cos\alpha \cos\beta + J_z \sin\alpha \sin\beta, \qquad (16)$$

$$J_x = \frac{1}{2\pi} M_1 M_2 \int e^{-ka}(1 - e^{-kh_1})(1 - e^{-kh_2}) \frac{k_x^2}{k} \xi(\mathbf{k}) \xi(-\mathbf{k}) dk_x dk_y \qquad (17)$$

$$J_z = \frac{1}{2\pi} M_1 M_2 \int e^{-ka}(1 - e^{-kh_1})(1 - e^{-kh_2}) k \xi(\mathbf{k}) \xi(-\mathbf{k}) dk_x dk_y \qquad (18)$$

Since $\xi(\mathbf{k})\xi(-\mathbf{k}) = |\xi(\mathbf{k})|^2$, both coefficients in Eq. (16) are positive $J_x, J_z > 0$. Note that the correlation of magnetic films interfaces plays a crucial role when performing integration in Eq. (1). The result Eq. (16) appears due to the correlation between the magnetization distribution in the second film and the magnetic field produced by the first layer. If the interfaces were not correlated, the OP effect would be zero.

The interaction strength depends on the thickness of the magnetic layers and on the thickness of the insulating spacer. If the thickness of the magnetic layers exceeds the characteristic in-plane length scale of roughness $d_{1,2} \gg L$ the interaction does not depend on $d_{1,2}$. This is quite clear since the stray field created by the roughness decays within the distance $L$ from the interface. When the magnetic layers are thin $d_{1,2} \ll L$ the interaction strength linearly grows with increasing of the layer thicknesses $J \sim d_1 d_2$. This is because the interaction volume grows with $d_{1,2}$ in this limit. Note that in most experimental magnetic multilayer systems of interest the film thickness can be rather small. Therefore, the limit of thin films is more relevant. The exponential decay of the "orange-peel" interaction on the insulator thickness is the same as was obtained in a seminal paper of Neel [1] and Ref. [5].

Equations (17) and (18) show that the magnetostatic interaction energy is anisotropic. The anisotropy means that the coupling strength depends on the angle between the layers magnetic moments and the average surface normal (z-axis in our case). Moreover, when the magnetizations are in the (x,y)-plane the interaction is ferromagnetic, while when $\mathbf{M}_{1,2}$ are along z-axis the interaction is antiferromagnetic. This result is in contrast to Ref. [5] where it was obtained that the OP interaction is absent (appears in higher order in $\xi_0/L$) in the case when $\mathbf{M}_{1,2}$ are along the z-axis. The reason for this discrepancy is the following. In the case of the in-plane magnetization the interface roughness leads to appearance of inhomogeneous surface magnetic charge linear in $\xi$. This charge produces the stray field of the same order. For the out-of-plane magnetization the inhomogeneous surface charge is of order $\xi^2$. It leads to vanishingly small stray filed which was calculated in Ref. [5]. In fact, there is another contribution to the stray field coming from the uniformly charged curved surface. Such a surface produces the stray field linear in $\xi$. The interaction strength in the case of the in-plane magnetization orientation is the same as in Ref. [5].

In the seminal paper of Neel [1] the check board height distribution $\xi(x,y) = \sigma \sin(qx) * \sin(qy)$ was considered in the case of the in-plane orientation of magnetization. Surface density of the magnetostatic interaction energy for such a roughness in the case of general orientation of magnetic moments is given by

$$\frac{W}{S} = \frac{\pi}{2\sqrt{2}} M_1 M_2 \sigma^2 q e^{-\sqrt{2}qa}\left(1 - e^{-\sqrt{2}qh_1}\right)\left(1 - e^{-\sqrt{2}qh_2}\right)(2\sin(\alpha)\sin(\beta) - \cos(\alpha)\cos(\beta)), \quad (19)$$

where $S$ is the surface area. In this case one has $J_z = 2J_x$. This relation holds for any surface with symmetry $|\xi(k_x, k_y)|^2 = |\xi(k_y, k_x)|^2$, for instance, for the Gaussian roughness with the following correlation function

$$<\xi(\boldsymbol{\rho})\xi(\boldsymbol{\rho}')> = \sigma^2 e^{\frac{-(\rho-\rho')^2}{2l^2}}, \quad (20)$$

$$<\xi(\boldsymbol{k})\xi(-\boldsymbol{k})> = 2\pi S \int \sigma^2 e^{\frac{-\rho^2}{2l^2}} J_0(k\varrho)\rho d\rho, \quad (21)$$

Where $l$ is correlation radius, $\sigma^2$ is the height dispersion, $J_0(k\varrho)$ is the zero-order Bessel function, Brackets $\langle ... \rangle$ denote averaging over the surface area.

In the consideration above we assumed that $\mathbf{M}_{1,2}$ are in the (x,z)-plane meaning that the in-plane components of these vectors are parallel. A more general expression can be written in the case of isotropic surface

$$W_{\text{op}} = -J_x M_{1z} M_{2z} + J_z (\mathbf{M}_{1\perp} \mathbf{M}_{2\perp}),$$

where $J_z = 2J_x$ and $\vec{M}_{1,2\perp}$ are the in-plane components of vectors $\mathbf{M}_{1,2}$.

To observe the angular dependence of the "orange-peel" effect one can use a ferromagnetic resonance method. It allows studying a magnetic interaction under an applied external magnetic field. Applying the external field along the sample plane one should observe the in-plane ferromagnetic interaction between magnetic layers. When the external field is applied perpendicular to the sample plane the out-of-plane antiferromagnetic interaction should be observed.

**Acknowledgements**

This research was supported by the Russian Science Foundation (Grant No. 16-12-10340).